\documentclass{emulateapj}
\usepackage{epsfig}
\usepackage{amsmath}

\slugcomment{Based on an analysis of six years of archival data from the AAOmega spectrograph at the Anglo-Australian Teleoscope.}

\shorttitle{Fibre-optic transmission stability}
\shortauthors{Sharp, Brough \& Cannon}

\begin{document}


\title{Long-term stability of fibre-optic transmission for multi-object spectroscopy}


\author{R. Sharp\altaffilmark{1}\email{rgs@mso.anu.edu.au}, S. Brough\altaffilmark{2}, and R.D. Cannon\altaffilmark{2}}
\affil{$^1$Research School of Astronomy \& Astrophysics, The Australian National University, Cotter Road, Weston Creek, ACT 2611, Australia\\
$^2$Australian Astronomical Observatory, PO Box 915, North Ryde, NSW 1670, Australia}

\begin{abstract}
We present an analysis of the long-term stability of fibre-optic transmission properties for fibre optics in astronomy. Data from six years of operation of the AAOmega multi-object spectrograph at the Anglo-Australian Telescope is presented. We find no evidence for significant degradation in the bulk transmission properties of the 38\,m optical fibre train. Significant losses ($\le20\%$ relative, 4\% absolute) are identified and associated with the end termination of the optical fibres in the focal plane. Improved monitoring and maintenance can rectify the majority of this performance degradation.
\end{abstract}



\section{Introduction}
Fibre-optics have long been used in astronomy as an expedient means for reformatting the focal plane for wide field-of-view multi-object spectrographs \citep{Hill88}. However, little is known about the long term transmission performance of such systems. This study addresses this stability via archival data from the AAOmega spectrograph \citep{Sharp06}. The data span a six year period commencing shortly after the initial commissioning of the AAOmega instrument at the Anglo-Australian Telescope.

\section{The 2dF fibre positioner}
The two-degree Field facility (2dF) at the Anglo-Australian Telescope (AAT) is a robotic positioning system for a multi-object fibre-optic fed optical spectrograph designed primarily with high-multiplex ($\sim$400 simultaneous targets) wide-area (2$^\circ$ diameter field in a single observation) galaxy redshift surveys in mind. The prime-focus optics and fibre positioning robot are described in detail by \citet{Lewis02}, here we give a brief review of the pertinent features of the fibre-optic feed.  The AAOmega spectrograph \citep{Saunders04,Smith04}, commissioned in November 2005, provided a significant upgrade to the 2dF facility, allowing increased observational sensitivity, an expanded range of observational modes/resolutions, and gravitational and thermal stability. However, locating the AAOmega system in the west coud\'e room on the fourth floor of the AAT dome required a fibre-optic cable some 38\,m in length. A decision was taken early in the life of the AAOmega project to maintain a single fibre cable without employing a demountable fibre coupling/connector stage.  This implies that the periodic removal of 2dF from the AAT requires a full removal of the fibre cable, with the associated handling and storage requirements for the delicate fibres. A key concern throughout the program has been the level of transmission degradation induced in the long fibre run due to gravitational loads and manipulation of the cable.

\begin{figure}
\begin{center}
\epsfig{file=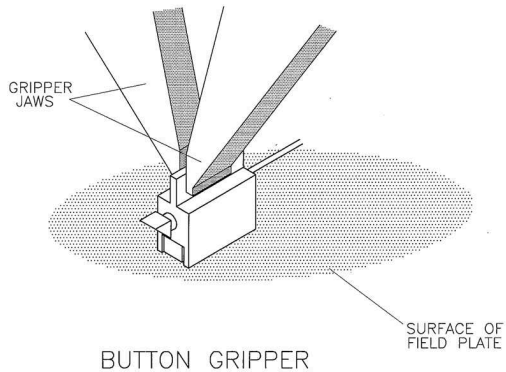}
\end{center}
\caption{\label{button} A schematic of the 2dF robotic fibre positioner gripper jaws and fibre button is shown. See text for a brief overview and \citet{Lewis02} for a detailed description.}
\end{figure}

A schematic of the 2dF robotic fibre positioner gripper jaws and fibre button is shown in Figure~\ref{button}. Fibres adhere to the steel field plate via a rare-earth magnet housed in a steel {\it button} (providing a sticking force of $\sim$50\,g). The button has a vertical fin which is grasped by the 2dF robot for repositioning. Total internal reflection within a glass prism (of approximately 90$^\circ$ angle) relays the target star/galaxy light, which arrives perpendicular to the {\it field plate surface} at the focal plane, into the 140\,$\mu$m diameter core ($\sim$2\farcs1) optical fibre. The fibre exits the back of the button parallel to the field plate before reaching the fibre retraction system.  Mounting of the fibre within the magnetic button proves to be critical to performance.  Laboratory tests show that point stresses of the fibre must be minimised to avoid light loss, either via direct loss through cladding leakage or more indirect losses within the spectrograph due to induced Focal Ratio Degradation (see below).  Efforts to minimise fibre stress are in conflict with the need to firmly secure the fibre into the magnetic button. The original strategy adopted for AAOmega
is as follows: each fibre is inserted into a steel ferrule and glued in place with UV setting cement.  The cement is fluid prior to UV setting and hence capillary action draws the cement along the full length of the fibre/ferrule bond. The fibre and ferrule are then optically polished to provide a smooth optical surface. In a similar manner the steel ferrule is in turn glued into the magnetic button housing. Finally, with the aid of a precision alignment jig, the 90$^\circ$ prism is bonded to the fibre/ferrule surface with the optically transparent UV cement. The fibre then runs free from behind the magnetic button, around a pair of 60\,mm diameter pulley wheels within the fibre retractor mechanism, before reaching a clamping point at the end of the retractor approximately 1.5\,m behind the fibre button.

\section{Light loss mechanisms}
A number of mechanisms exist which could result in an apparent reduction in the transmission properties of optical fibres with time.

\vspace{0.25cm}
\noindent{\bf External loss:} A decrease in the overall transmission of the telescope or spectrograph optics will manifestly reduce transmission of the fibre system. The AAT primary mirror is typically re-aluminised in February each year, with $\sim$10\% annual variations in reflectivity not uncommon.  Preliminary analysis of the AAOmega collimator reflectivity indicates a spot reflectivity measured at $\sim$95\% of that expected based on the specification, suggesting a measurable global degradation at this surface over the lifetime of AAOmega. Unfortunately, no records of the reflectivity of the mirror on commissioning are available to confirm this.

\vspace{0.25cm}
\noindent{\bf Fibre Fringing:} Small air-gaps between two optical surfaces lead to a classic \'etalon interference effect. AAOmega fibres appear to be susceptible to the formation of such an air-gap between the prism and the fibre face \citep{Sharp06}. In this mode, the optical prism remains firmly mounted to the steel ferrule but becomes detached from the optical fibre surface due to differential thermal expansion between the glass fibre and the steel ferrule.  The \'etalon imprints a significant amplitude (up to 10\%) modulation in the fibre transmission with wavelength.  Small air gaps (10\,$\mu$m) induce $\sim$20 cycles across the 3000-9000\,\AA\ spectral range of the AAOmega system.
A number of fibres exhibit this fringing effect, but with long term stability, allowing the feature to be removed with accurate flat fielding.  However, this does not overcome the implicit transmission loss at the destructive interference positions.  A smaller fraction of fibres show unstable fringing patterns (on the time scale of typical observations) which cannot be removed with a flat field and which cripple sky subtraction for such fibres \citep{Sharp10b}.  Experimentation indicates the instability in these fibres is triggered when repositioning the fibre, likely via stresses induced in the fibre button, with relaxation times of 2--3\,hours. Ensuring a perfect bond between the fibre and the prism, via careful matching of thermal expansion coefficients of all critical components, has been shown to mitigate much of the fringing problem for AAOmega.
 
\vspace{0.25cm}
\noindent{\bf Focal Ratio Degradation:} The multi-mode optical fibres typically used in astronomy accept light at an initial input angle (or $f$-ratio from the telescope) and eject it with a wider cone angle (lower focal ratio) due to internal scattering processes as the light propagates along the fibre (violating the simplistic {\it light pipe} model typically envisaged as their mode of operation). This phenomenon is known as Focal Ratio Degradation \citep[FRD;][]{Ramsey88} and its implications are considered by \citet{Parry06}. The practical upshot of FRD for AAOmega is degraded instrumental resolution and the necessity for the spectrograph to accommodate faster angles than those at the focal plane. Any increase in FRD beyond that accommodated in the spectrograph design will result in light overfilling the spectrograph optics and being lost to scattering within the spectrograph. Simple laboratory experimentation easily demonstrates that stressed fibres (e.g., an inappropriate clamp point or a tight bending radius) suffer significant FRD.  \citet{Haynes11} also demonstrate that poor fibre end preparation will result in significant modal scattering and FRD.  There is clear scope within the AAOmega fibre button construction and the fibre retraction mechanism for significant FRD generation, particularly as the fibres are repeatedly reconfigured in the focal plane over time.  Measurement of FRD in optical fibres is complex, and near impossible in the operational AAOmega fibre system, and as such little is currently known about the level of FRD in the AAOmega fibre run.

\vspace{0.25cm}
\noindent{\bf Micro-fractures:} Microscopic structural damage to the fibre cable, induced by continued exercising of the fibre retraction system, by incautious handling, or simply by gravitational loading along fibre while mounted on the telescope would reduce bulk transmission (and likely increase FRD).\\

Identification of the scale and origin of light-losses within the AAOmega fibre feed would inform significant design improvement for future fibre spectroscopy systems. Any assessment of the long term performance of the optical fibres must take into account any maintenance processes undertaken during operations.

\subsection{Fibre repairs}
\label{repair}
While continued improvements to the 2dF hardware have greatly reduced the rate of fibre breakage during routine operations, there are a number of remaining failure modes which result in broken fibres requiring repair. The first is a failure of the retraction mechanism intended to withdraw an individual fibre as it is parked or positioned in the focal plane by the 2dF fibre positioning robot. This typically leads to fibres looping on the focal plane and around obstacles (e.g., other fibre buttons, the retractor mechanisms or the 2dF robot). This inevitably leads to sharp kinks in the fibre, snapping it typically a few inches behind the fibre button (where the fibre is rigidly mounted to the button). The second failure mode is caused by the 2dF robot {\it stamping} fibre buttons onto previously placed fibres. This latter mode is typically the result of previously failed fibre retraction leaving fibre loops on the focal plane. After 15 years of operations fibre collisions due to failure of the software that models fibre placement interactions are rare. A final major mode results when prisms detach from individual buttons due to positioning errors or during an overly enthusiastic manual intervention. 

Most failure modes result in a fibre break within the last 10--100\,mm of fibre behind the magnetic button. The majority of fibre repair work consists of trimming the fibre to remove the damaged section, re-terminating and polishing the fibre face, and replacing the fibre magnetic button.
For instances of missing prisms, the fibre/ferrule can be re-polished within the
button and a new prism bonded to the fibre without trimming.
In all cases, a repaired fibre continues to use much of the fibre length that has previously been mounted within the fibre retraction mechanism. The fibre run does contain a limited spare-length loop (1\,m) but a fibre that was to break repeatedly would soon become unserviceable. Additionally, $\sim$100 spare fibres were included in construction of the AAOmega fibre run in 2005. These were intended primarily to guard against incidents during the construction phase of the project. They were not required and have not been deployed with the 2dF positioner.

The result of the repair process is that fibre repairs will not recover any degradation caused within the majority of the 38\,m length of the AAOmega fibre run, nor due to wear within the 2dF top end fibre retraction system. They will only overcome catastrophic breakage at the fibre button and progressive degeneration at the fibre button, such as shrinkage
at the polished surface inducing {\it fibre fringing}.\footnote{Experiments with a small number of 2dF retractors indicate that employing a glass ferrule to house the fibre can eliminate differential expansion and suppress formation of the interference fringing. While glass ferrules pose significant polishing problems due to their reduced structural integrity, this has been overcome for the upgraded 2dF feed to be deployed for the AAOmega/HERMES system.}

\section{Data format and processing}
Long term archival data sets are used in this analysis (see \S\ref{surveys}). Due to the data volume involved, data have not been reprocessed, and hence the original survey program data reductions have been used. The 2dF
fibre-feed to the AAOmega spectrograph consists of two interchangeable sets of 400 optical fibres. In any given observational configuration, eight multi-core fibre bundles are used for acquisition and guiding leaving 392 optical fibres to feed the input slit of the AAOmega spectrograph, giving a total of 784 science fibres.

Of the 392 fibres in a given observation, 20--30 are allocated to blank sky positions to facilitate sky subtraction. At any given time the fibre system may operate with 5--10\% of fibres disabled due to attrition during observations, with fibre repair work scheduled at convenient intervals. Fibres are arranged in banks of ten, held within individual mechanical fibre retraction stages. Practical limitations mean it is highly inefficient to repair a single fibre within an individual retractor block and so the repair interval for a single isolated fibres can be significant in some cases.

Spectra have been processed from the raw CCD frames via the conventional range of analysis steps. In all cases the data are prepared using the \texttt{2dfdr} data reduction package developed by the AAO for 2dF/AAOmega, although this has evolved significantly over the lifetime of the survey programs used in this analysis. The basic extraction process is outlined in \citet{Sharp10a} while the source of many of the imperfections in processed spectra are discussed in \citet{Sharp10b}.

A combination of data from two extragalactic survey programs provides an archival data set spanning almost seven years of operation with the AAOmega fibre cable, sampled monthly with the exception of the first six months of operation before the first of the surveys commenced.

\subsection{Archival data}
\label{surveys}
Archival reduced data has been provided by the WiggleZ \citep{Drinkwater10} and GAMA \citep{Driver11} long-term galaxy surveys undertaken at the AAT with AAOmega.
The WiggleZ Dark Energy project observations obtained spectra of faint emission-line galaxies (targeting [O\textsc{ii}]\,$\lambda$3727 over the interval $0.5 < z < 1.4$) with low continuum signal-to-noise ratio spectra (S/N~$\leq$~1 per pixel, for galaxies with $r$(AB)~$<$~22.5), while the GAMA galaxy survey records modest S/N spectra ($r$(AB)~$<$~19.8, S/N~$\geq$~5 per pixel) suitable for galaxy redshift determination via cross-correlation analysis.
Neither source class significantly degrades the measurement of the relative strengths of the OH sky-lines imprinted onto galaxy spectra and which are to be used for the analysis presented in \S\ref{transmission}.
The GAMA project spectra span the wavelength interval 5700-8600\,\AA, while the WiggleZ spectra were recorded in two distinct wavelength modes, 5700--8600\AA\ and 6700--9600\AA, at different stages of the project. In the analysis that follows, only data from the red arm of the dual beam spectra, where the night-sky signal is dominate by the OH airglow lines, are used.

These data sets provide a number of opportunities to monitor the longterm stability of the optical fibre transmission. In \S\ref{abscal} we investigate two approaches that use the available survey data to perform a direct calibration of long term transmission variations. These methods are however subject to variations in external conditions (such as seeing changes and cloud cover). In \S\ref{transmission} we develop a relative transmission measure, which is free from these uncertainties.

\section{Absolute flux calibration and long-term sensitivity variations}
\label{abscal}
The archival datasets used in this analysis all include calibration stars within each fibre configuration. These stars are typically white-dwarfs (WD) and hot sub-dwarfs (SD) due to their largely featureless smooth spectra which are well modelled by a black body. They are included to provide a contemporary assessment of flux calibration for each observation. While in principle these internal calibration stars provide an opportunity to perform an absolute flux calibration, and hence transmission analysis of the survey data, this is not feasible in practice. An accurate absolute flux calibration is not possible from the fibre data alone due to the unknown aperture losses from: the coupling of atmospheric seeing to the fibre diameter; acquisition effects (centring stars repeatably on a fibre); and overall atmospheric transmission variations \citep[see][for a detailed review of these effects]{Newman02}. 
Rather the stars are included largely to confirm the internal relative spectral shape calibration for individual observations, monitoring field-to-field variations during a survey due to changing atmospheric conditions.

However, modulo a significant scatter in the individual measurements, one would expect the mean absolute calibration derived from these sources to be constant with time.

\subsection{Standard star observations}
There are some 1268 calibrator stars (including repeat observations) in the 3.5 year WiggleZ survey data set (typically 2--3 per WiggleZ survey field) and 2969 observed as part of GAMA. We use these measurements to assess the long-term stability of the AAOmega system.

\begin{figure}
\epsfig{file=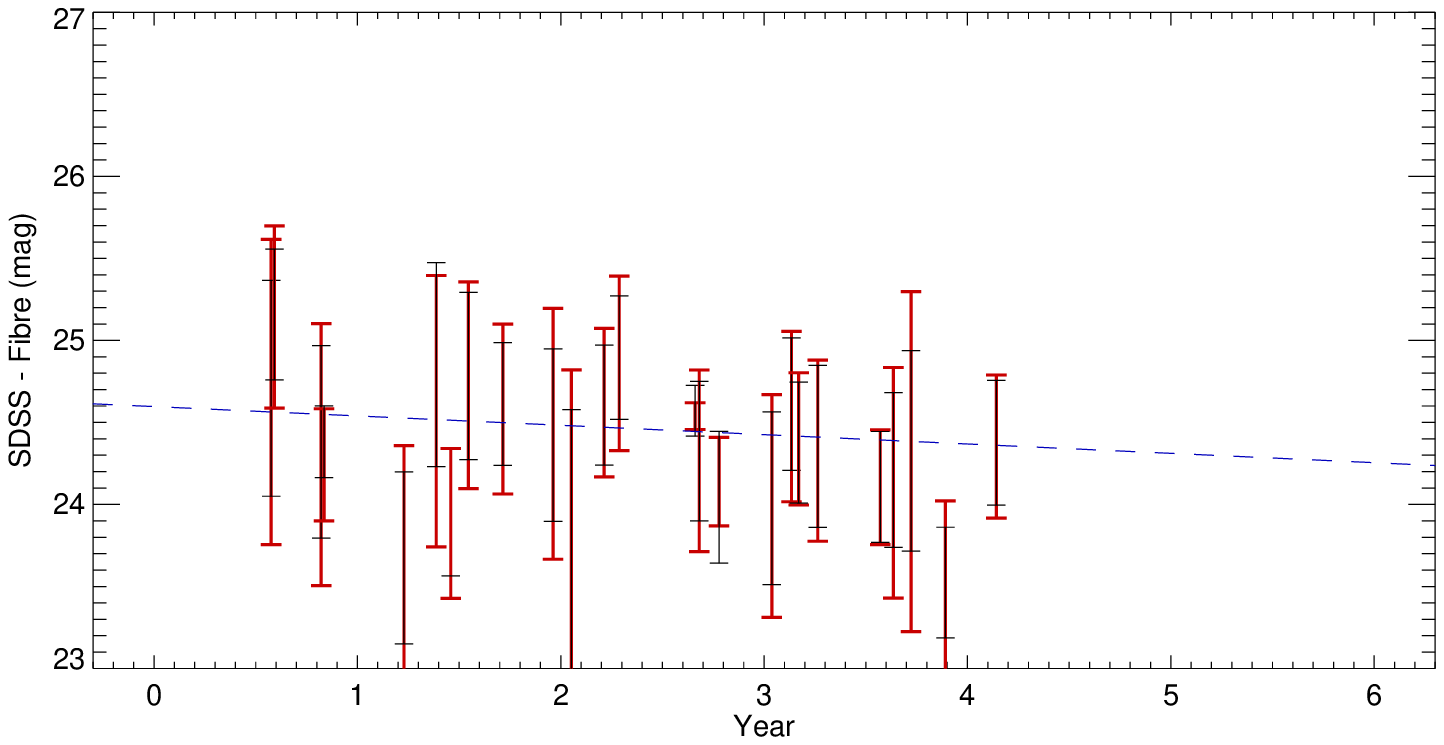,width=8cm}
\epsfig{file=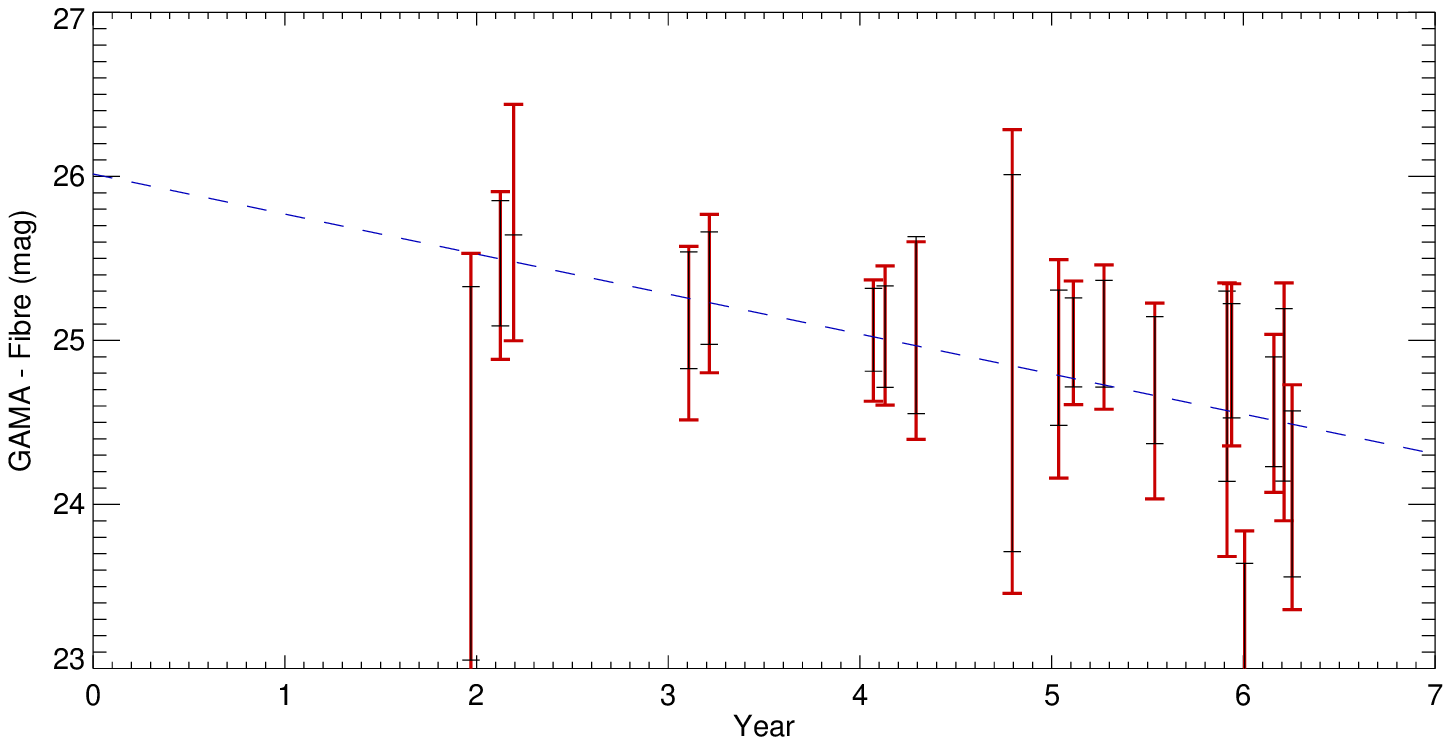,width=8cm}
\caption{\label{WiggleZstd} The evolution of the spectroscopic zero-point can be derived from standard star measurements taken during survey observations. The time-averaged results, with a significant scatter due to the modest number of data points and highly variable observing conditions (see text for details), are show for the WiggleZ (upper panel) and GAMA (lower panel) surveys. Error bars show the 1-standard deviation and inter-quartile range for each data bin. There is evidence for a decline in sensitivity from the two survey data sets.}
\end{figure}

A fibre magnitude is derived from each calibration star spectrum by expressing the median count rate in the spectrum as a magnitude, mag(fibre)=$-$2.5$\times$log$_{10}$($<$count rate$>$). The count rate is taken from the wavelength range that corresponds to the SDSS $r$-filter.\footnote{The SDSS $r$-filter is approximated here as a square filter with $\lambda_{\rm eff}$ = 6231\,\AA; FWHM = 1373\,\AA}
This quantity is then compared to the known SDSS $r$(AB) magnitude for the calibration source in question. While the effective zero-point of the fibre magnitude system is of limited value, its long-term stability is a measure of the consistency of the fibre transmission properties over the lifetime of the survey.
The calibration results are shown in Figure~\ref{WiggleZstd} with the data sampled at monthly intervals to overcome the significant scatter in individual source measurements  due to variations in external factors. Based on a simple linear fit to the zero-point data for the WiggleZ survey, there is evidence for a $\times$0.8 decline in mean fibre transmission over the 3.5\,year span of the survey although with large uncertainties consistent with no variation. A precipitous decline is in principle seen in the equivalent GAMA survey data, however the decline is dominated by excellent seeing conditions for observations in year 2 and substandard fibre positioning performance compounded by poor weather in year 6. This highlights the difficulty of obtaining a direct measure of absolute transmission.

\subsection{GAMA variations in redshift completeness}
\label{Gama}
\begin{figure}
\begin{center}
\epsfig{file=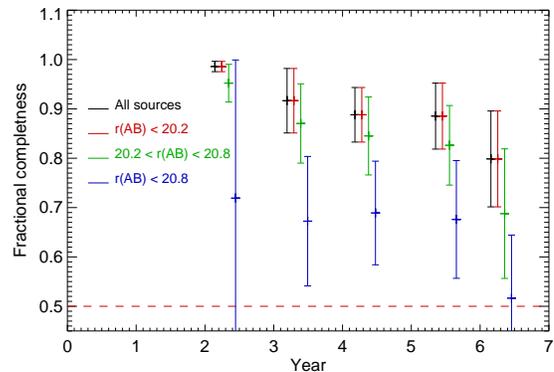,width=8cm}
\end{center}
\caption{\label{GAMAcomp} The reported redshift completeness (percentage of high quality redshifts recovered per observation) for the GAMA survey as a function of target input magnitude. The SDSS $r$(AB) 3\arcsec\ diameter $fibremag$ is used (2dF fibres are $\sim$2\farcs1 diameter).  The median and one standard deviation values are shown for 560 GAMA survey fields. Thirty-six fields with completeness below 50\% are removed under the assumption of incomplete observation or peculiarly adverse weather. Small horizontal offsets are applied to the points for clarity.}
\end{figure}

An alternative analysis is provided by an apparent drop in the survey completeness for the GAMA survey. Figure~\ref{GAMAcomp} presents an analysis of the percentage of GAMA sources targeted for which a {\it high quality} \citep{Driver11} redshift is returned.  The data are broken up into bins based on the reported SDSS $r$(AB) 3\arcsec\ diameter $fibremag$ (2dF fibres are $\sim$2\farcs1 diameter).  There is a strong apparent decline in the survey completeness in later years, particularly for fainter targets more sensitive to small transmission variations. It should be noted that the seeing during the first year of observation was exceptional for the AAT ($<1.2$\arcsec\ as measured through the 2dF prime focus corrector at Zenith Distance, ZD $\sim$ 30$^\circ$) for 21.5 of the 22 nights of observation.  While observations in year 6--7 (Jan--April 2012) have been compromised by weather, this has largely been through telescope closure as opposed to obviously poor atmospheric transmission.  There is evidence that recent mechanical modifications to the 2dF robot (replacement of the linear motors is known to have changed the differential distortion mapping) have introduced a small but significant ($\sim$0\farcs3) increased positioning error in fibre placement which affected data between November 2011 and Febuary 2012 \citep[see][for a discussion of the consequence of such a misalignment]{Newman02}. The GAMA survey commenced $\sim$2\,years after the commissioning of AAOmega, leading to the lack of data to provide this comparison in years 1--2.  The dynamic selection boundaries of the WiggleZ survey make those data unsuitable for a similar analysis.

\section{Throughput estimation from all fibres}
\label{transmission}
A fibre relative transmission estimator is derived for each individual observation in order to facilitate sky subtraction for the faint object science fibre spectra \citep{Sharp10b}.
Briefly, of the $\sim$400 fibres available in a given observational configuration with AAOmega, of the order 20--30 are allocated to blank sky positions. On data reduction, the blank sky spectra are combined to provide a master sky spectrum, essentially free from Poisson noise due to the combination of a number of independent spectra. Before combination, however, the individual spectra must be scaled and normalised. This normalisation is undertaken by measuring a constant spectral feature prominent in all blank sky (and also source) spectra; the OH airglow emission lines \citep[see][for a recent review of OH airglow properties]{Ellis08}. The integrated flux of a set of well resolved OH lines is measured for all spectra (both for blank sky and science targets) and, under the assumption that the sky presents a uniform intensity across all fibres in the 2\,degree FoV during an integration (typically of the order 1200--2400\,sec), the relative transmission of each fibre is normalised according to the integrated OH line flux.

Due to the inherent wide variations in the OH intensity during any given night's observations, as well as variations in other observational conditions, this normalisation provides only a relative measure of transmission across the $\sim$400 fibres in a given observational data set.
However, if there is no temporal variation in the transmission properties of the fibre set, then the distribution of relative transmission values will remain constant over time, scattered only by the limitation in stability of individual measurements.

For the available processed survey data, the relative transmission vector for all fibres in each data frame was recovered. Information on the fibre assignment (S--Sky, P--Program, U--Unavailable [i.e., disabled] or N--Not-used [i.e.,  not allocated]) was also extracted for each individual fibre, as was the date of observation and the plate number for the fibre (the two 2dF field plates are populated with 800 physically distinct fibres, 400 per plate). Fibres classified as unavailable or unused are retained in the reduced data (to maintain a stable data format), and while this data is of course meaningless on a frame-by-frame basis it is valuable for rapidly flagging periods during which a fibre was disabled before fibre repair work.

\begin{figure*}
\epsfig{file=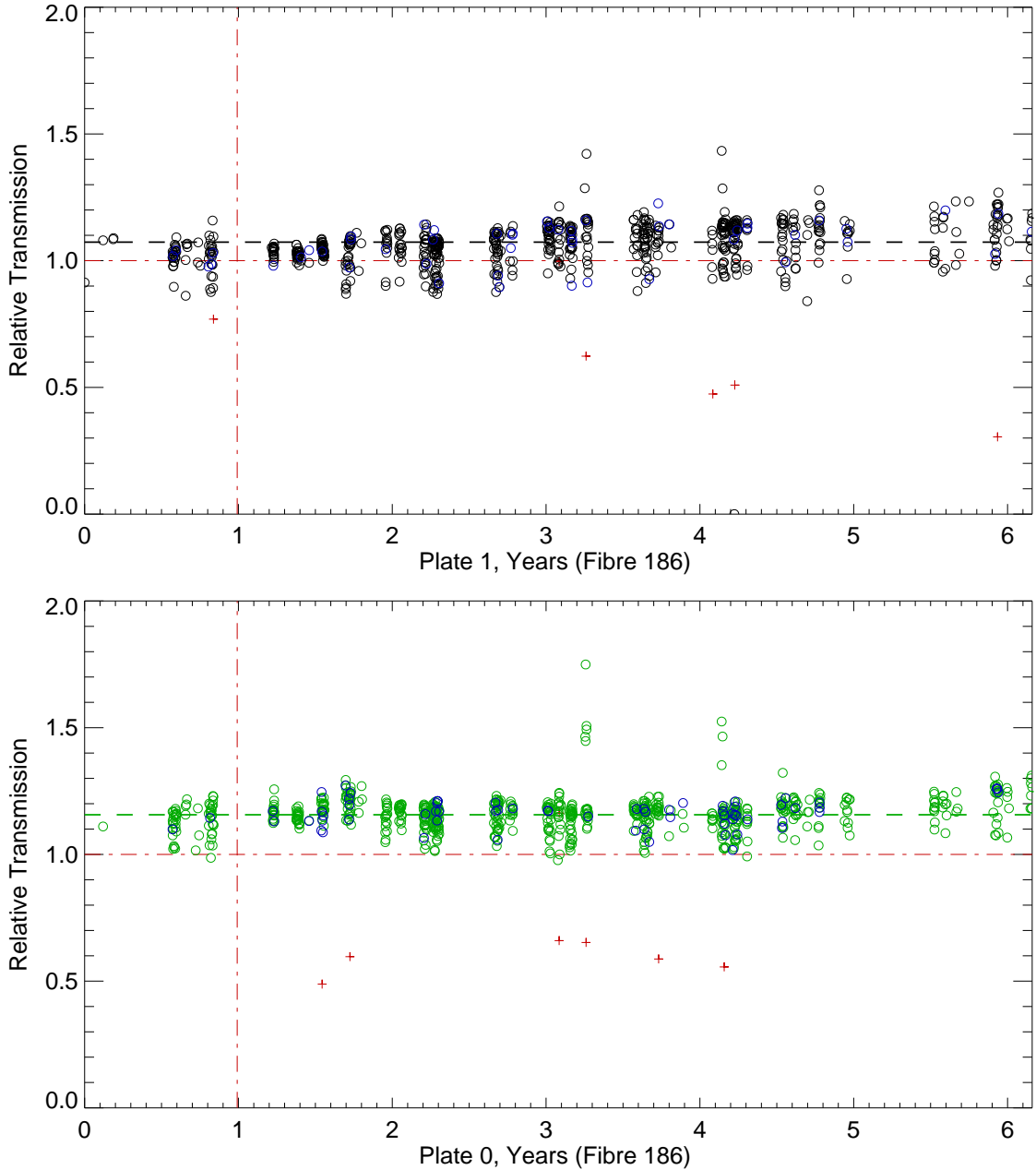,width=15cm}
\caption{\label{goodfib} The lifetime transmission variations are shown for two stable Fibres (Fibre 186 on Field plate 0 and plate 1). Both fibres start their lives with above average relative transmission, and remain uniformly high across the data set. Black/Green circles show transmission variations estimated from science spectra, Blue circles show those from sky spectra. Red symbols show unused or unallocated spectra, and are included to show the complete fibre history. Horizontal lines show the median sample transmission (unity by design) and the median transmission for the single fibres in turn. The vertical line indicates the only intervention for these fibres - the in-situ re-glueing of all AAOmega fibres into their magnetic buttons.}
\end{figure*}

\begin{figure*}
\epsfig{file=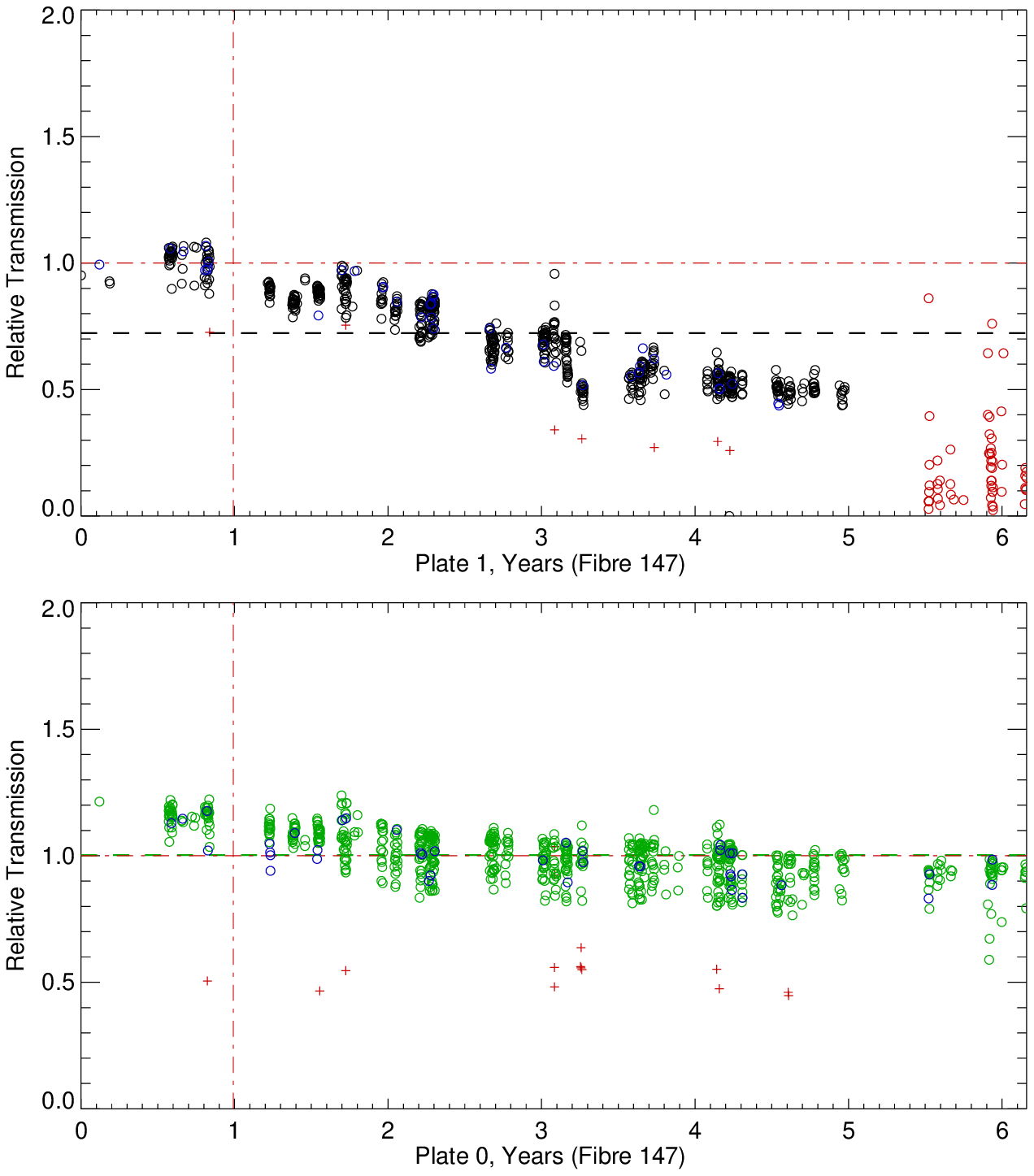,width=15cm}
\caption{\label{badfib} Figure~\ref{goodfib} if repeated for two {\it poor} fibres, number 147 on field plate 0 and plate 1.  Both show a significant decline in transmission with time, before an undocumented catastrophic event removed the plate 1 fibre from operation.}
\end{figure*}

The relative transmission properties of two fibres which show uniform response over their operational lives are shown in Figure~\ref{goodfib}, two less fortunate fibres are shown in Figure~\ref{badfib}. The two uniform fibres (Figure~\ref{goodfib}) have been in operation since November 2005. All AAOmega fibres required intervention in January 2006 when it was discovered that the glue used to set fibres into the magnetic buttons was failing on a significant fraction of fibres and all fibres were re-glued {\it in situ}.

\subsection{Limitations of the relative transmission measure}
There are limitations to the use of the relative transmission measures as a metric for the time variation of the transmission properties of the AAOmega fibres. While the relative transmission measures are perfect for uniformly illuminated fibres (as is the case for the sky background) and facilitate excellent sky subtraction accuracies
\citep{Sharp10b}, they are unreliable for absolute sensitivity calibration for a number of reasons.

Firstly, there is significant scatter in the estimator for any given fibre, as is clearly visible in Figure~\ref{goodfib}. Some of this scatter is intrinsic to the observational system. For example radial distortion from the prime focus corrector introduces a plate scale change across the 2\,degree FoV, and is considered below.

Secondly, and more importantly, long-term secular gradients in transmission will not be recovered by this metric since the median transmission of any given observation is, by definition, normalised to unity. 

Additionally, before the relative transmission data can be used to assess changes in the overall transmission properties of the AAOmega fibre-fed one must correct for any known causes of apparent transmission variation within the data. Failure to account for any known systematic variations would inflate the scatter in fibre transmission measures and dilute the transmission loss model (to be developed below) which is based on the broadening over time of the distribution of transmission measures. As we will see in the following section, approximately 30\% of the width of the distribution of relative transmission measures in data taken soon after installation of the fibre-run can be accounted for by known systematic effects.

\section{Radial variations in transmission}
\label{radial variation} 
The variation in relative transmission with the radial position of a fibre on the 2dF field plate is assessed for the full sample in Figure~\ref{radialVariations}. A strong radial structure is evident in the median relative transmission binned radially, with a peak value seen at around 1/3 of the way from the field centre falling off by 10\% at the edge of the field plate.
Vigneting is minimal within the 2$^\circ$ field \citep{Lewis02} but is seen markedly for fibres in the {\it parked} position and hence such fibres are trimmed from the analysis. A number of known effects introduce a change in the expected relative transmission of fibres as a function of field position, and these must be accounted for to avoid artificially inflating the scatter in the average observed transmission properties.

\subsection{Apparent fibre diameter}
The 2dF prime focus corrector has a significant radial distortion across the 2$^\circ$ field. This results in a plate-scale change with position and hence an apparent change of fibre diameter on the sky for the physically constant 140\,$\mu$m fibre cores. Since the relative transmission measure is normalised to the OH signal strength within each fibre, the increased sky flux from the enlarged aperture results in a higher apparent relative transmission. The plate scale varies from $2\farcs16$ at the field centre to $2\farcs0$ at the maximum extent of the 2$^\circ$ field
(the distortion terms being well defined as they are essential for the astrometric distortion model).  The 8\% radius variation leads to a 15\% flux variation with the radial position of the fibre on the field plate.
The reduced effective area of a fibre when at large field plate radius dictates the form of the observed reduction in the relative transmission seen in Figure~\ref{radialVariations}. The significant rise at small plate radius is counteracted the by non-telecentric angle modulation discussed below.

\subsection{OH background variations}
The OH airglow is known to vary rapidly both temporally and spatially.  While the central limit theory indicates that many of these variations should average out over the course of an extended AAOmega observation, a global intensity gradient is expected to remain due to the geometric variation in the integrated OH column length as seen across the $\pm$1$^\circ$ field of 2dF \citep[as defined in equation 9 of][based on the earlier work of \citet{Content96}]{Ellis08}.  This residual is predicted to be of the order $\sim$1\% across the 2dF field-of-view, aligned with the paralactic angle, for observations at ZD~$\sim$~30$^\circ$.

\subsection{Focal Ratio Degradation variations}
The focal ratio of the 2dF corrector varies roughly linearly with radial position. This modulates losses due to Focal Ratio Degradation with plate position. The slower beam for fibres at the edge of the field plate results in a smaller collimator angle for these fibres in the spectrograph, and hence reduces the impact of light losses from overfilling the spectrograph collimator which would be the natural consequence of serious FRD \citep{Parry06}.

\subsection{Non-telecentric variations}
The 2dF prime focus corrector delivers light to the prime focus at an $f$-ratio~=~3.4 on axis and perpendicular  to the field plate.   At the edge of the field plate, the incoming beam angle is inclined at 4$^\circ$ to the normal, introducing a significant light loss due to non-telecentricity.  This loss was partially compensated in the original 2dF design \citep{Lewis02} through the use of 92$^\circ$ prisms, which correct the non-telecentric beam variation to $\pm$2$^\circ$ across the field plate.  Additionally, the prime focus corrector also presents a slower $f$-ratio at the edge of the field ($f$-ratio=3.7) which, in principle, would allow a reduction in light-loss due to the changing field angle, by biasing the prism angle towards minimising the non-telecentric  angle in the centre of the field plate, where the cone angle is wider due to the faster $f$-ratio.  However, since the majority of the plate area is located at the edge of the field plate, a prism angle of 92.5\,deg$^\circ$ was selected for AAOmega.  This biases the feed angle to higher radial positions, maximising the global transmission over the field, at the expense of moderate additional losses ($\sim$5\%) for the inner most fibres).

\subsection{Correcting for radial variations}
The top panel of Figure~\ref{radialVariations} presents the observed radial variations in the sample. The contributions to this distribution from the variable plate-scale (dashed green curve) and from non-telecentriciy plus FRD (dot-dashed blue curve) are shown in the middle panel. The true form of the FRD loss is poorly constrained at this time and hence only a representative model has been used here.
A low order polynomial fit to the observed radially binned data is used to provide a plate-position dependant multiplicative correction to the relative transmission of each fibre in the sample. This correction flattens the median relative transmission, removing the signature of the radial variation and accounting for the apparent fibre aperture variations and the effects of non-telecentric illumination of the fibres. 

\subsection{Retractor based modulation}
\begin{figure}
\epsfig{file=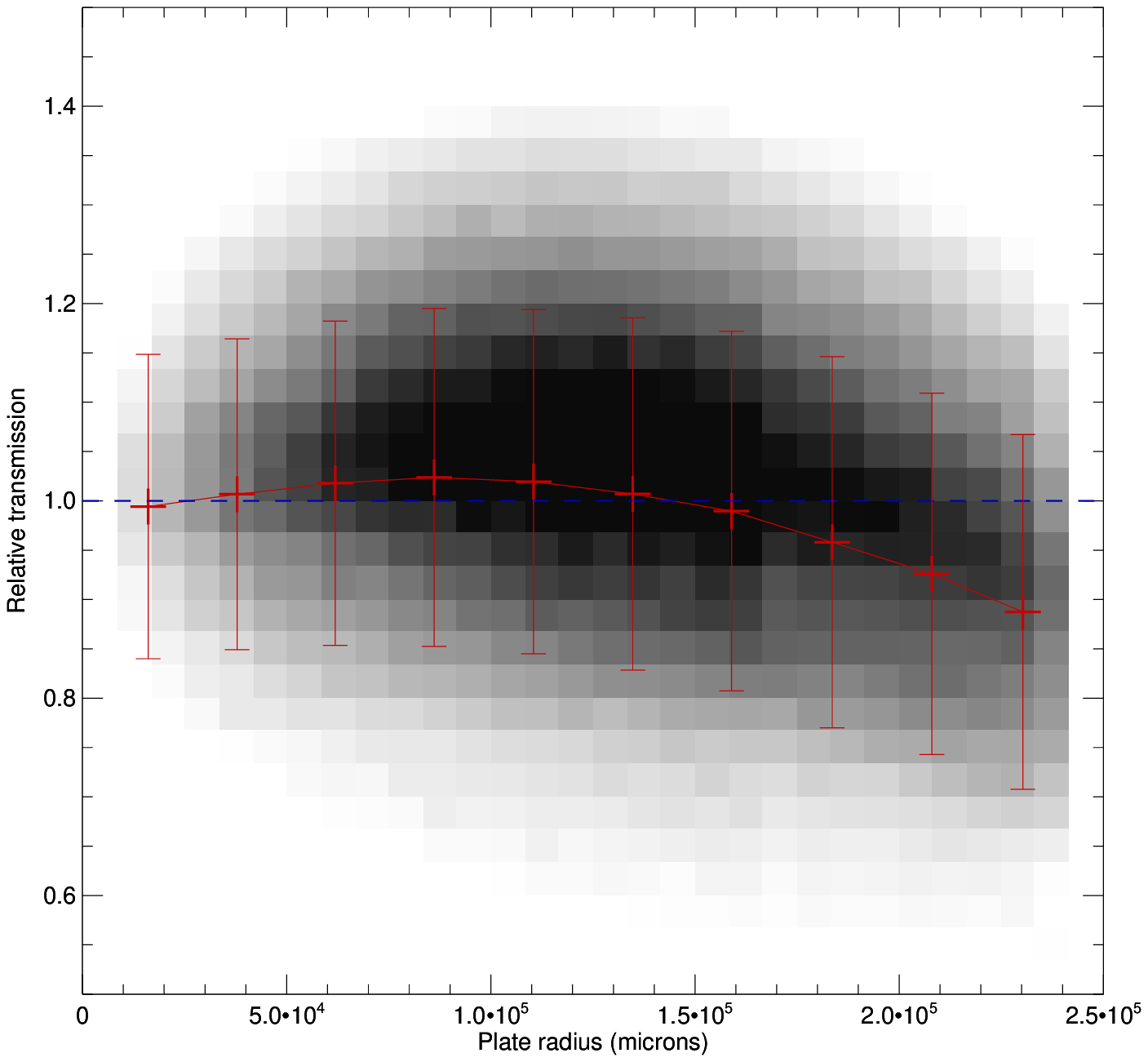,width=7.5cm}\\
\epsfig{file=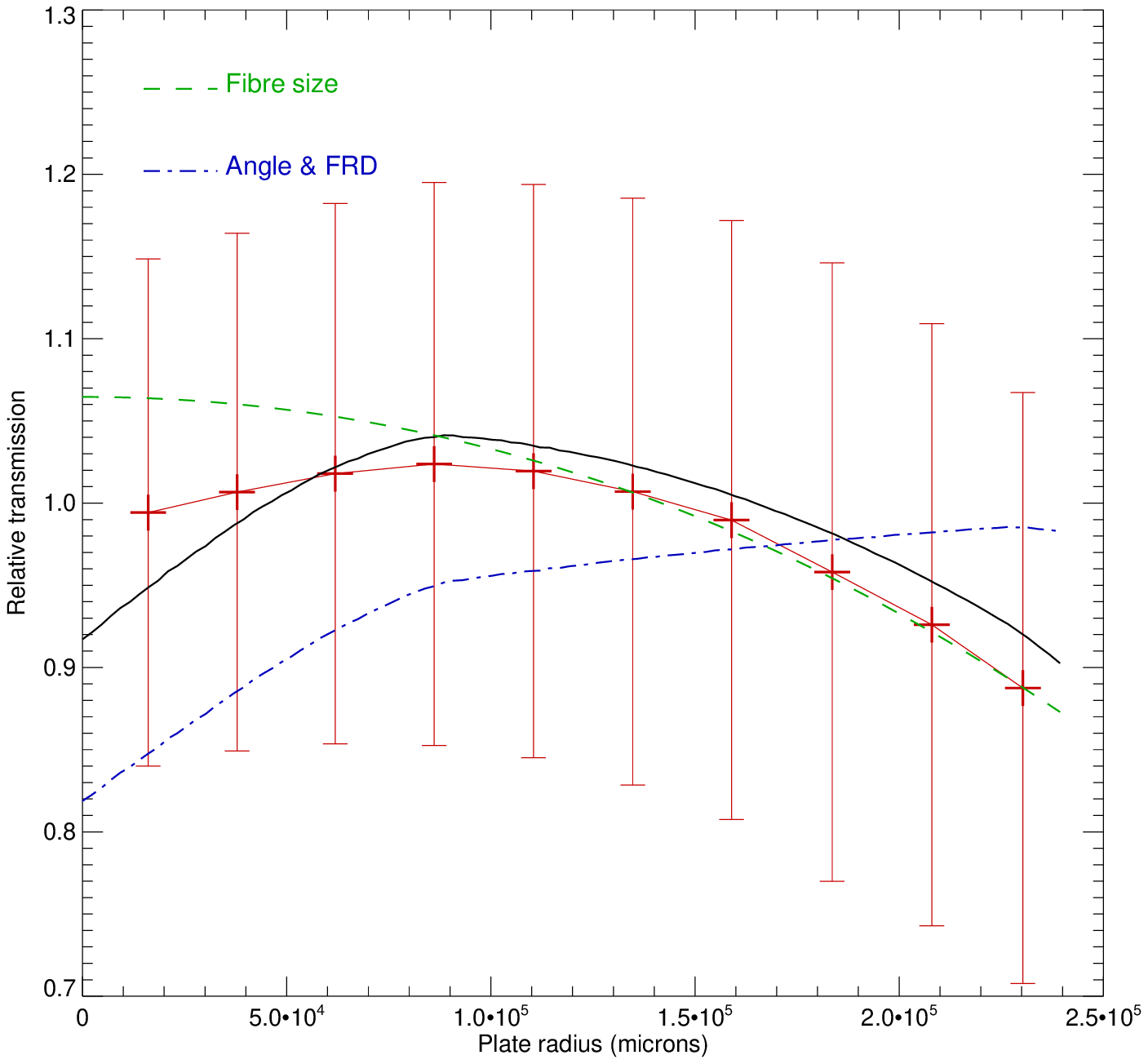,width=7.5cm}\\
\epsfig{file=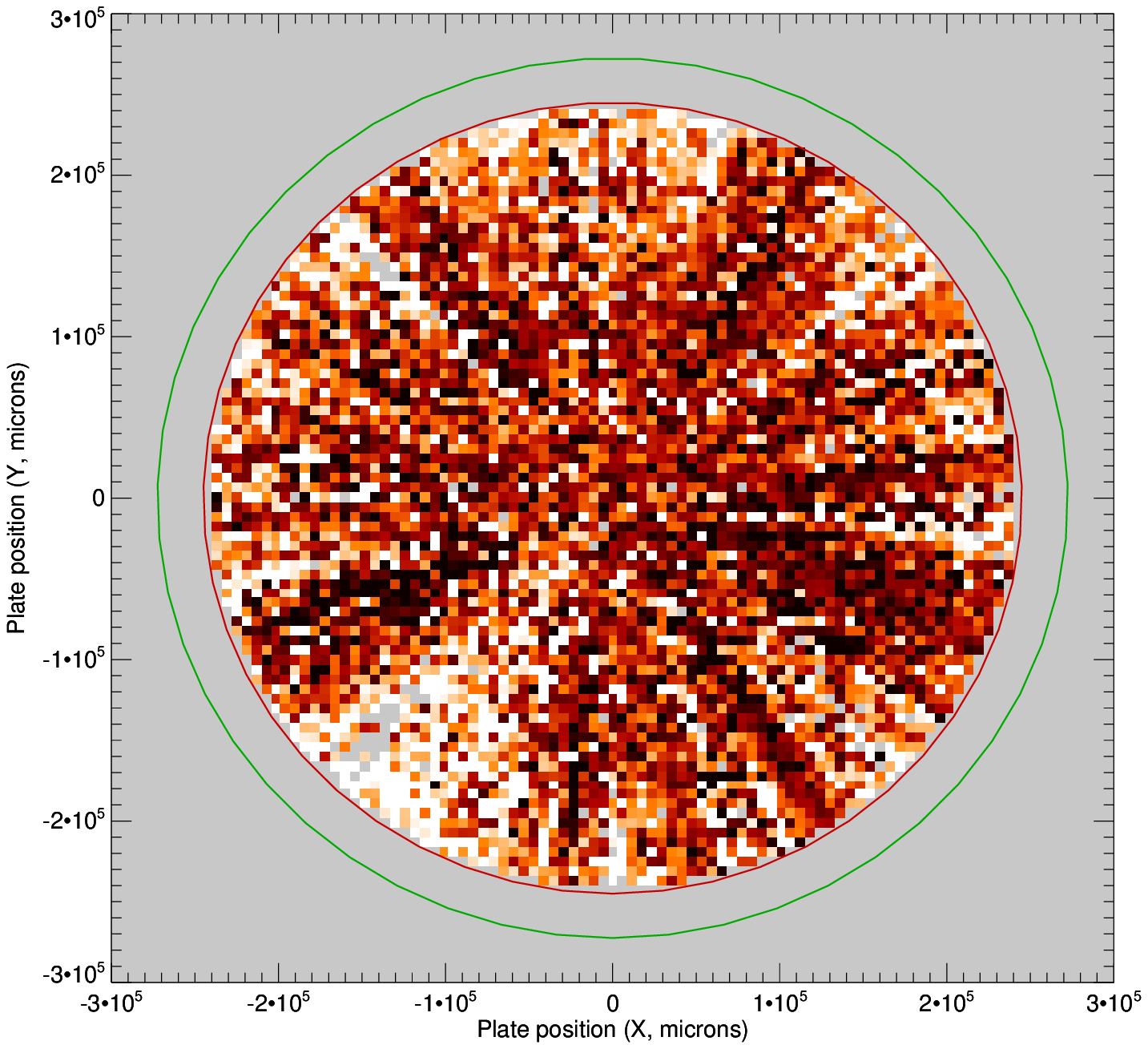,width=7.5cm}
\caption{\label{radialVariations} The radial variation in relative transmission is shown for all fibres in the sample. The upper grayscale shows a 2D histogram of the full fibre sample (both field plates) transmission distribution with a radially binned median $\pm$1 standard deviation in each bin indicated.  The central panel indicates the relative contributions of variable fibre aperture size and none-telecentric light losses (with a basic FRD model included). The angular modulation around the field plate is shown in the lower image (white/black saturation is set to 0.8~$<$~T$_{rel}$~$<$~1.2). The data is truncated to that from active science fibres on the 2dF field plate (red circle), with the heavily vignetted {\it parked} fibres omitted (outer green circle).}
\end{figure}

\begin{figure*}
\epsfig{file=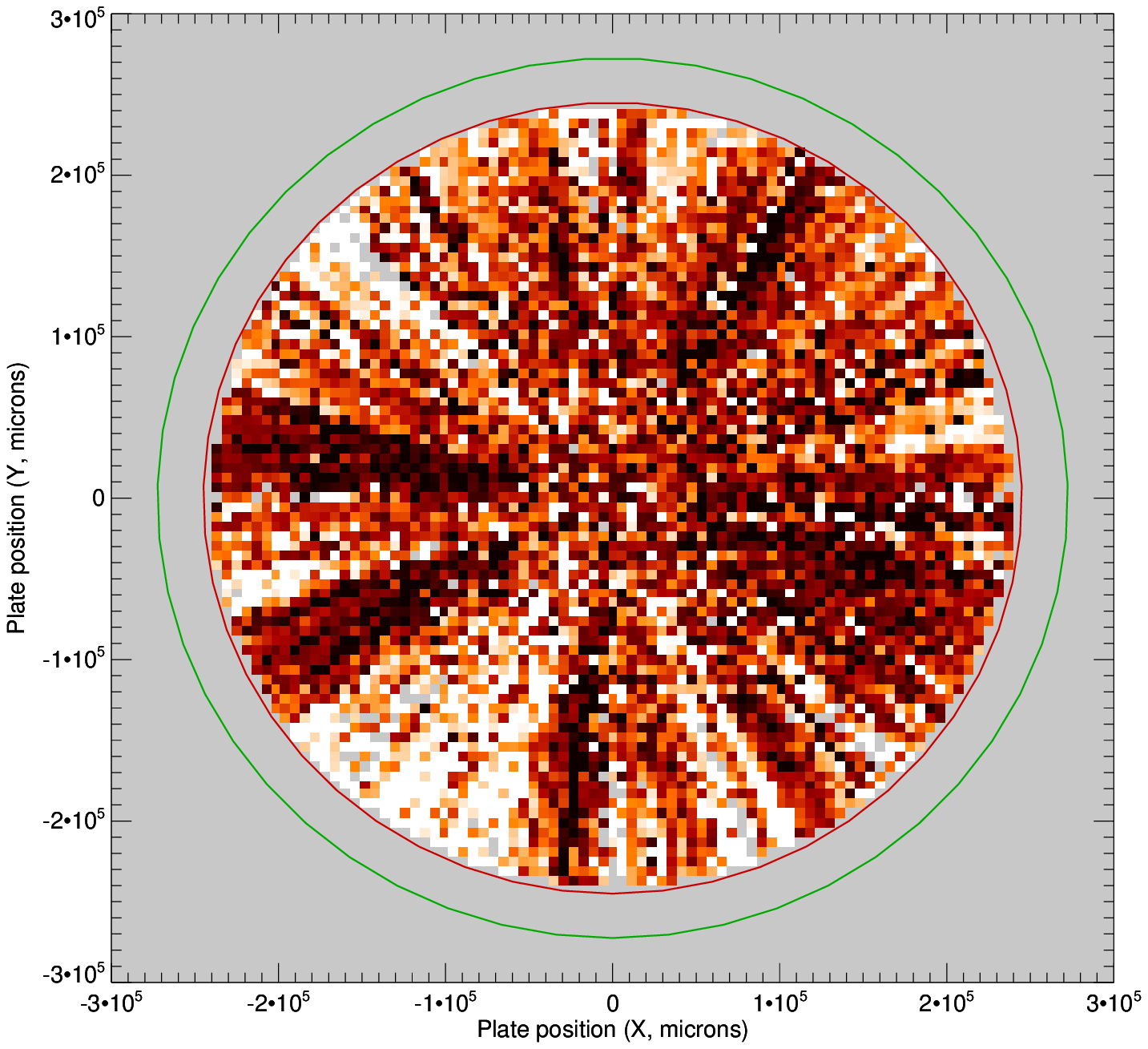,width=9cm}
\epsfig{file=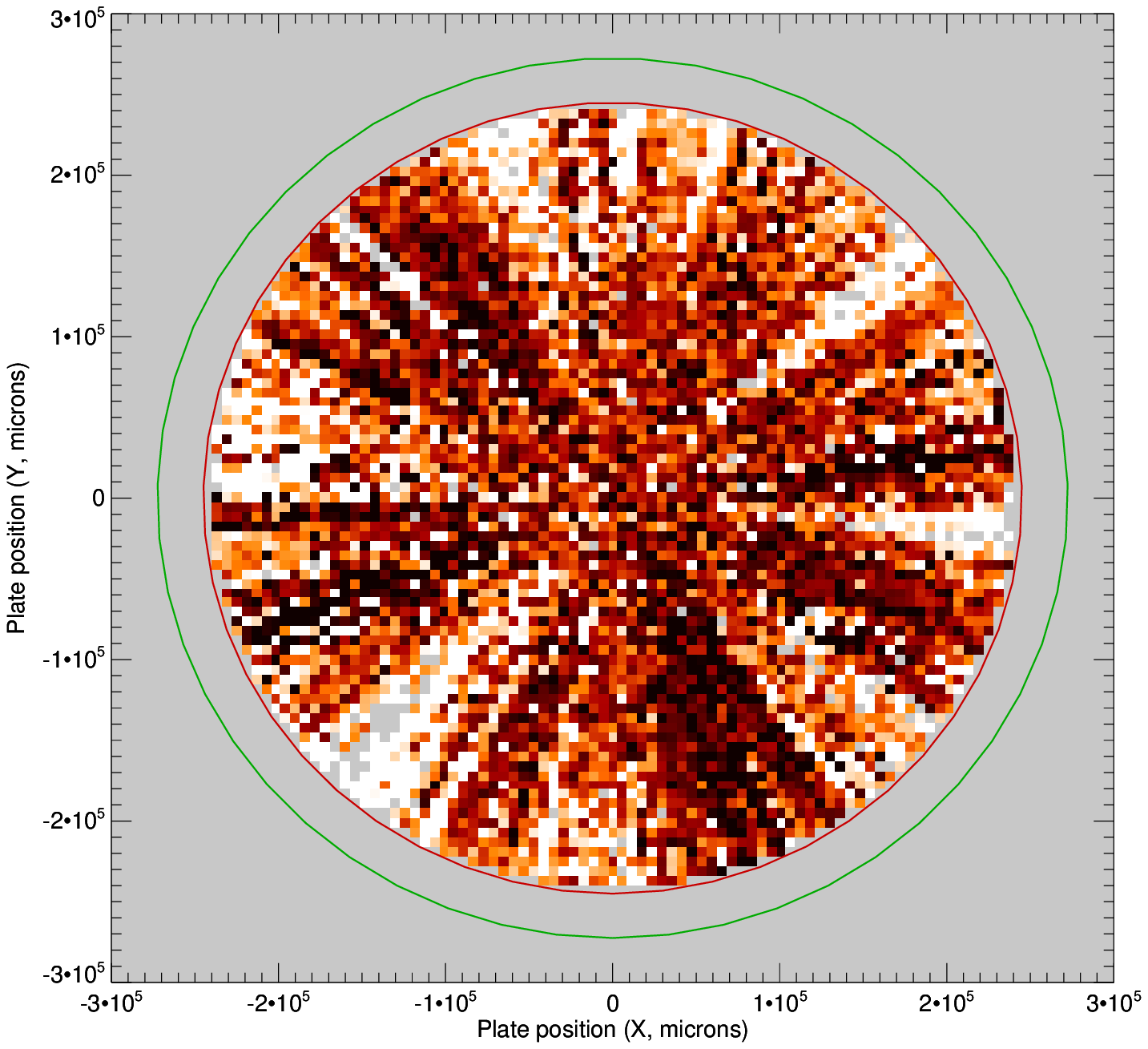,width=9cm}
\caption{\label{radialVariationsPlate} When the radial distribution of Figure~\ref{radialVariations} is reproduced for the two field plates independently, a strong retractor-by-retractor modulation of the relative transmission is clear. The grayscale is chosen here to saturate at 0.8~$<$~T$_{rel}$~$<$~1.2.}
\end{figure*}

The 400 fibres of each of the two field plates are arranged around the circumference of each plate and housed in forty independent retractor banks, each housing the fibre retraction mechanisms for ten fibres.
The lower panel of Figure~\ref{radialVariations} and Figure~\ref{radialVariationsPlate} highlight a stark contrast in relative intensities across individual retractor banks. This indicates that the major loss mechanisms are associated with manufacturing and assembly defects within the last meter of fibre (i.e., within individual retractors) rather than the bulk properties of fibres within the cable run. Fibre construction is typically undertaken in blocks of a full retractor, and hence all fibres within a retractor will be subject to the same environmental variations (e.g., temperature, glue batch, specific assembly jig, polishing materials, operating practices of the individual technicians). Changes in suppliers of components such as the magnetic buttons, optical prisms, or mounting ferrules, over the instrument lifetime will impact individual fibre repairs (although as noted earlier, fibre repair priority is typically given to retractors with multiple damaged fibres).

\section{Long-term sample variation}
\begin{figure*}
\epsfig{file=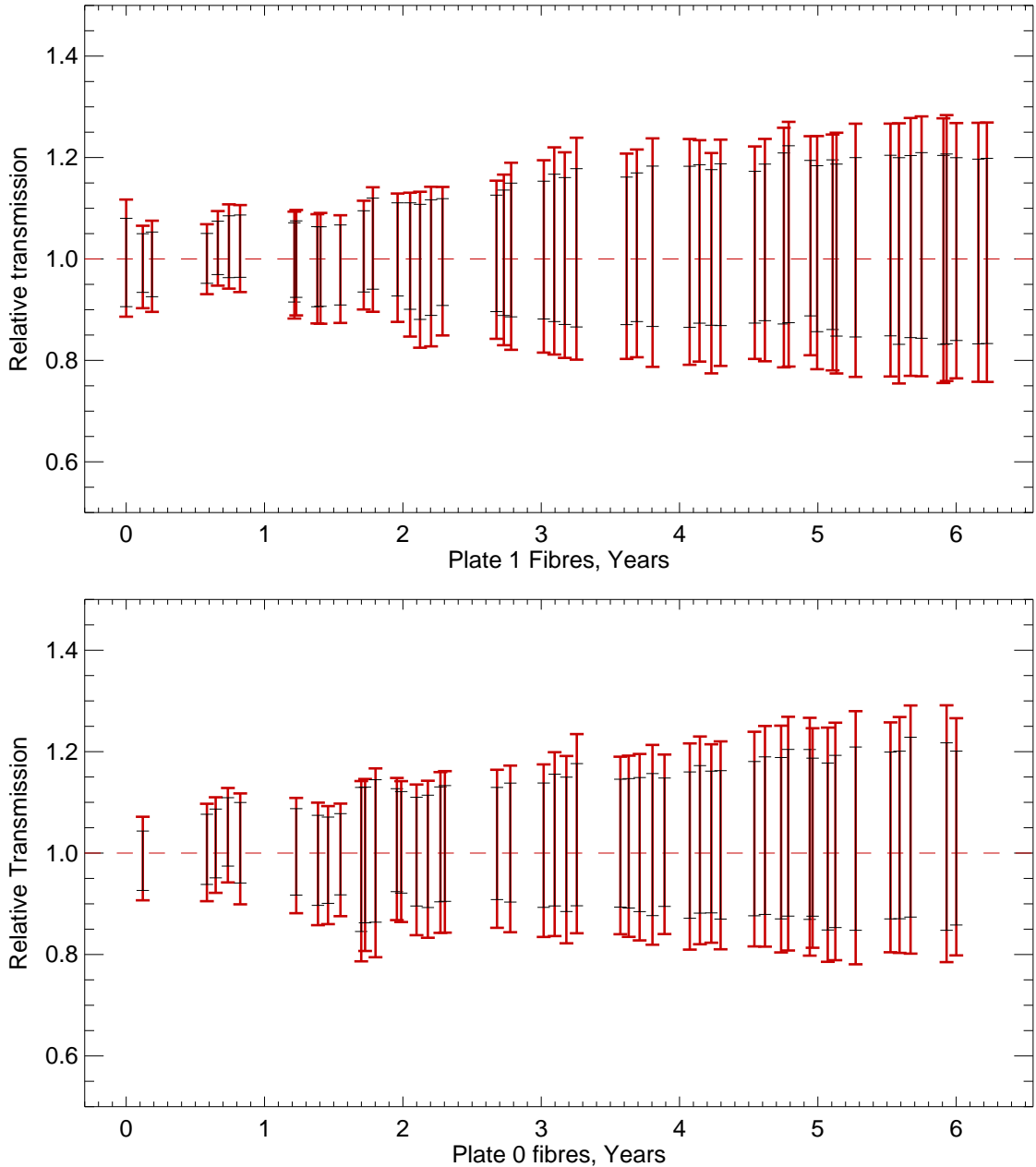,width=15cm}
\caption{\label{AllRelTran} Fibre relative transmission histories for all $\sim$500,000 science fibres from the archival data set are shown. The fibre transmissions are binned into one month wide samples and the median transmission is calculated for each bin, along with the bin standard deviation (red) and interquartile-range (black). By definition the median relative transmission in each bin is unity.  The data sets from Plate 0 and Plate 1 are analysed separately to provide two independent traces, although the result is largely the same. The radial variation fit from Figure~\ref{radialVariations} has been removed from the sample data prior to binning.}
\end{figure*}

Long-term variation in the sample relative transmission distribution is shown in Figure~\ref{AllRelTran} for the full six and a half years of WiggleZ and GAMA survey data, and for all $\sim$800 2dF/AAOmega fibres in each observation.

\begin{figure*}
\epsfig{file=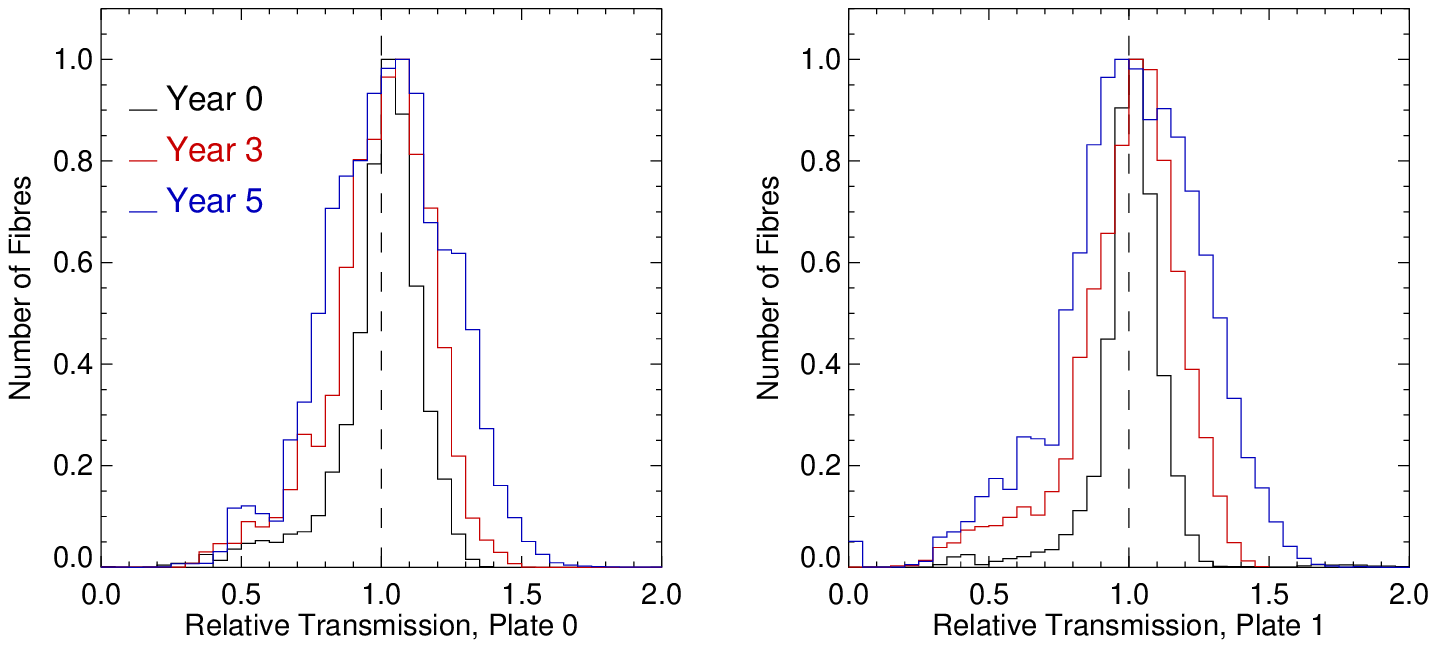,width=15cm}
\caption{\label{AllRelTranHist} Histogram distributions are shown for three intervals summed over a year's observation (i.e., cross sections of the fibre mean transmission histories of figure~\ref{AllRelTran}). While the mean relative transmission remains unity by definition, the distribution becomes significantly wider with time. The known radial variation fit from Figure~\ref{radialVariations} has been removed from the sample, while a significant component due to intrinsic sky intensity variations across the 2dF field remains.}
\end{figure*}

The fibre transmissions are binned into one month intervals and the median transmission is calculated for each interval, along with the standard deviation (red) and interquartile range (black). By definition the median relative transmission in each bin is unity.  The data sets from 2dF Plate 0 and Plate 1 are presented separately to provide two independent traces, although the result is largely identical.  There are transmission estimates for 266,193 and 262,400 fibres on the two plates respectively (reflecting the availability of science and sky spectra over the six year time scale) . There is a considerable increase in the scatter with time. Not all months contain survey observing runs and hence there are gaps in the coverage.  The sample size in each bin is variable, although sufficiently large in each case (typically 1,000--10,000 fibres) for statistical sample variance to be largely irrelevant within each interval bin.  Figure~\ref{AllRelTranHist} compares transmission histograms in intervals of one year width. A small number of early bins are missing for plate 0 due to a formatting problem in the early data reduction software which was incorrectly recording data for one field plate. There is unambiguous evidence for an increased scatter in transmission values with time, and in the absence of any known mechanism which would improve a fibres transmission over time, the scatter is attributed to a general decline in the mean fibre transmission of the sample. In principle, large-scale systematic replacement of fibres could induce improvements in the median transmission (see \S\ref{repair}).  However, no such systematic global fibre replacement has been undertaken to date. A new fibre-feed will be installed in late-2012 in preparation for the arrival of the HERMES high-resolution spectrograph to be commissioned with 2dF in 2013.

Global improvements to telescope operations (primarily positioner accuracy and target acquisition), data reduction processes or weather-related transmission effects would affect sky observations for all fibres equally and hence would not affect the median relative transmission or scatter.

\section{A basic model for transmission degradation}
\label{modelsection}
\begin{figure}
\epsfig{file=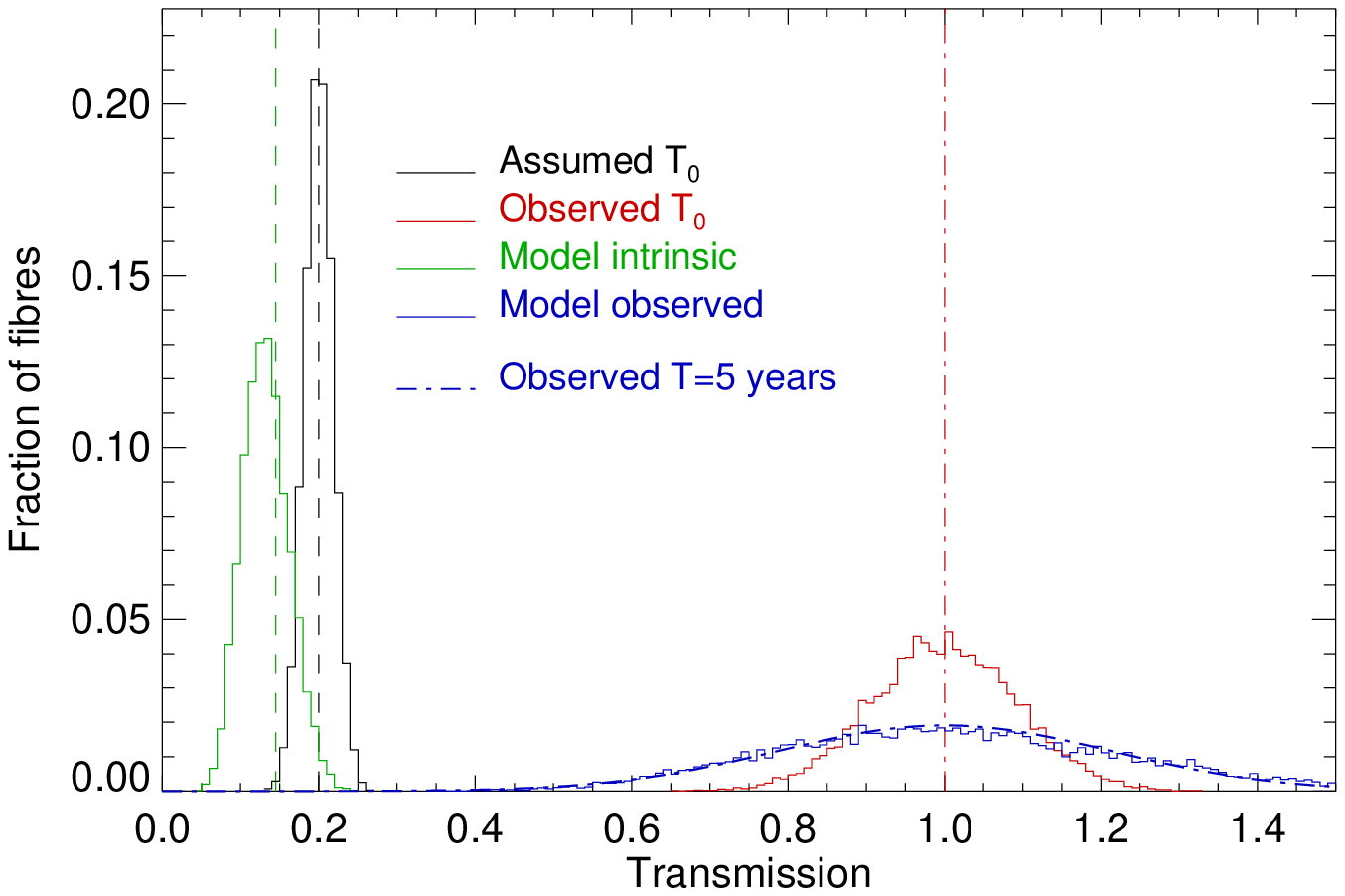,width=8.5cm}
\caption{\label{Model} A simple model of the evolution of the fibre transmission is shown. The black histogram shows an estimate of the intrinsic fibre transmission at time $\tau = 0$. The red histogram represents the resulting observed relative transmission distribution. The green histogram is of the intrinsic fibre transmission after application of a simple stochastic transmission loss model. The blue curve is the evolved observed relative transmission distribution, matching that of the year five interval from Figure~\ref{AllRelTranHist}. A significant reduction in the median transmission of the system is required to reproduce the observations.}
\end{figure}

Figure~\ref{Model} shows a simple stochastic model of a systematic reduction of fibre transmission with time.
The model assumes an intrinsic scatter in the transmission distribution of the newly installed fibres with a normal distributed centred on the intrinsic mean transmission of the system \citep[measured at 20\%:][]{Sharp06}. This width is estimated from the scatter in transmission data observed in early data from Figure~\ref{AllRelTranHist}, after correcting in quadrature the mean scatter observed for repeated measurement of individual fibres.

This initial transmission distribution is then evolved by stochastically introducing transmission reduction in individual fibres over successive iterations. This process broadens the normalised fibre transmission histogram as show in Figure~\ref{Model}. 
The black curve presents the assumed intrinsic model distribution for the fibre set at installation, with the observed mean for the AAOmega red arm.  The red histogram is the observed model after applying the measurement error inferred above and then normalising to a relative transmission of unity.
The green histogram shows the result of the simple evolutionary model after the transmission has decayed sufficiently that the the associated relative transmission histogram (solid blue curve) matches that observed in the 5 year AAOmega data from Figure~\ref{AllRelTranHist}. 

The conclusion from this simplistic model is that a significant decrease in mean fibre transmission is necessary to reproduce the observed increase in scatter in the normalised transmission data seen in Figure~\ref{AllRelTranHist}. While a global transmission loss will broaden the distribution due to its effect on the normalisation of the observed relative transmission, the stochastic process is preferred as a global loss of the magnitude required to produce the observed broadening is inconsistent with the modest performance losses reported for AAOmega earlier. However, the simple stochastic model most likely overestimates the median transmission loss for the fibre distributions, since by its nature few fibres exhibit extreme transmission variations and no fibre has a transmission {\it improvement} over time. The model can be informed via an analysis of the individual fibre transmission histories.

\section{Individual Fibre life histories}
\label{classification}
In order to assess the validity of the simple model generated in \S\ref{modelsection} we investigate the life histories of individual fibres as recorded in the AAOmega fibre maintenance database. The database records key events such as fibre repair or cleaning.  Repairs include replacement of the prism with or without re-termination of the fibre in a new steel ferrule. In any repair process that reveals the bare optical fibre, a re-polishing is always performed. In January 2006 (year~1 of operation) a significant number of fibres where found to be misaligned to the sky, with the fibre and prism rotated off-axis in the magnetic button.  At this time, all fibres were re-glued into their respective buttons to prevent rotation. The problem was traced to a poor choice of glue in the initial construction - the cyano-cryolite adhesive used was replaced with UV cement. This year~1 intervention is marked in the fibre transmission history plots by a vertical line.

\begin{figure}
\epsfig{file=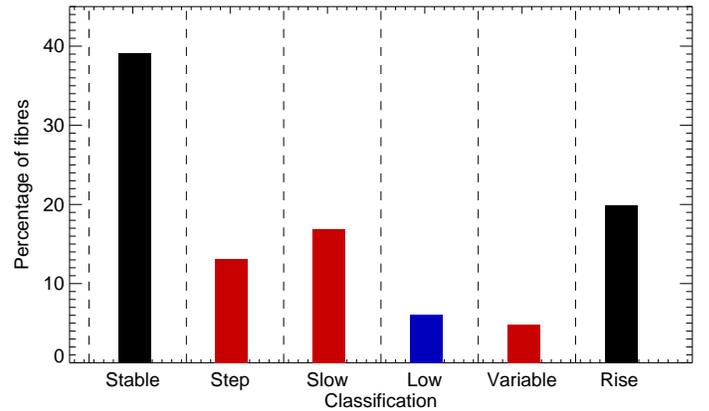,width=9cm}
\caption{\label{FibreClass} The fibre classification distributions from \S\ref{classification} are shown graphically. The Stable and Rising classifications represent fibres with {\it good} relative transmission data, while the Step, Slow and Variable classifications show performance degradation over time. The low classification represents poor fibre performance typically associated with a specific quality control failure during manufacturing (e.g., prism damage or gross miss-alignment).}
\end{figure}

The life histories for the AAOmega fibre set, from plate 0 and plate 1 were visually classified via the following classification scheme. The percentage of fibres that fall within each classification is shown graphically on Figure~\ref{FibreClass}.

\begin{itemize}
\item Stable -- consistent and moderate scatter around a stable relative transmission over extended time intervals without recorded intervention. Fibres with small (typically $<$10\%) step variations in transmission (up or down) associated with specific repair work are included while long-term rises in relative transmission are classified below.
\item Step-decline -- a step reduction in the relative transmission occurring over the course of less than a single observing block (i.e., instantaneous in the available data set).
\item Slow-decline -- a slow decline in the relative transmission, either as an ongoing decline or to a new stable value with $>$20\%  reduction over a time frame in excess of 1--2\, months. Fibres which begin a slow decline, before or after an extended period of stability, are included in this category.
\item Stable-low -- A stable relative transmission, over an intervention free period of greater than a year, but at a relative transmission below 90\%.
\item Variable -- a highly variable transmission, likely linked to multiple manual interventions. Typically this represents transitions between long periods of stability after repair work.
\item Rising -- a fibre whose relative transmission rises steadily over an extended period of time (typically six months to the full six year baseline). Almost all examples are for a relative transmission rise from an initial value of 0.8-1.0 to a final value of 1.0-1.3.  This high final relative transmission for a small number of fibres, is of great concern since, without a clear explanation for a transmission improvement over time it is assumed these fibres simply decline more slowly than  the majority of fibres.
\end{itemize}

\begin{figure}[h]
\epsfig{file=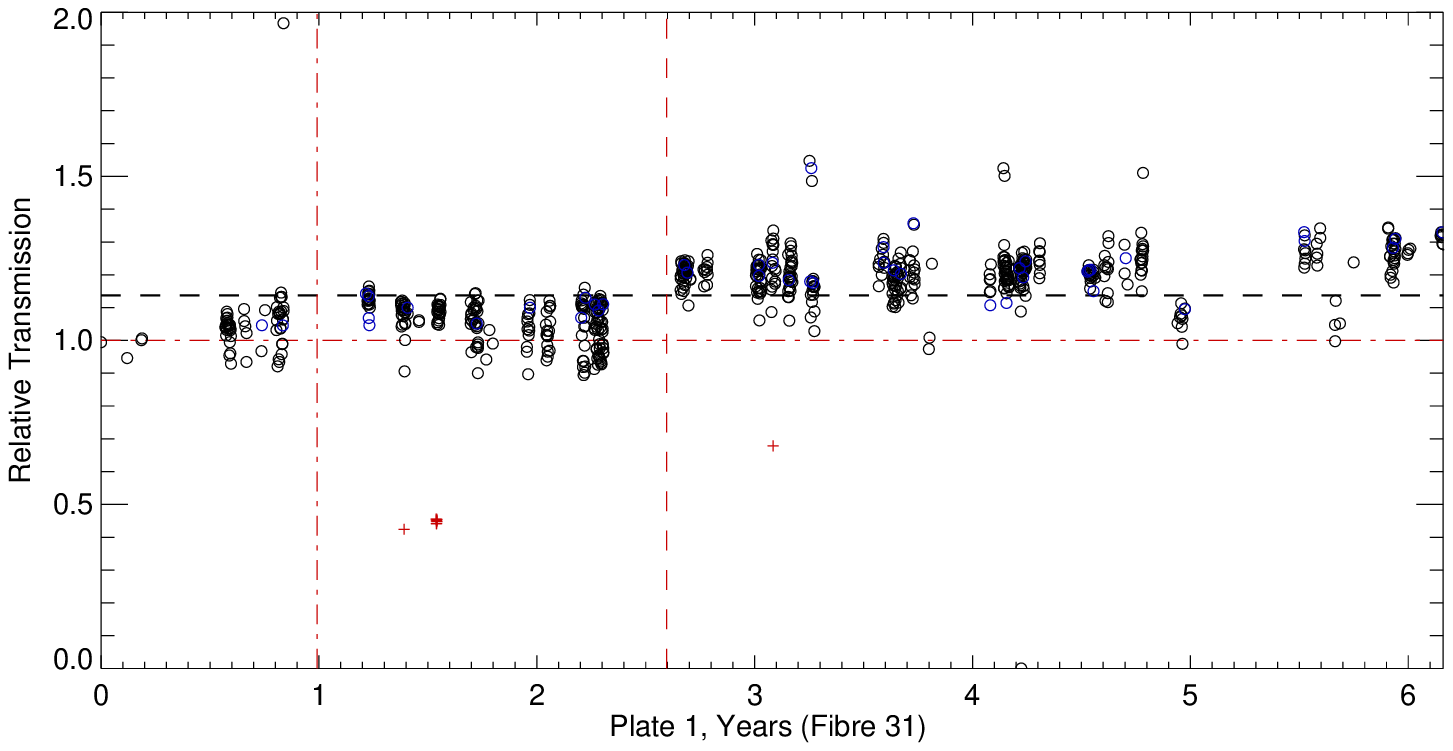,width=8cm}
\epsfig{file=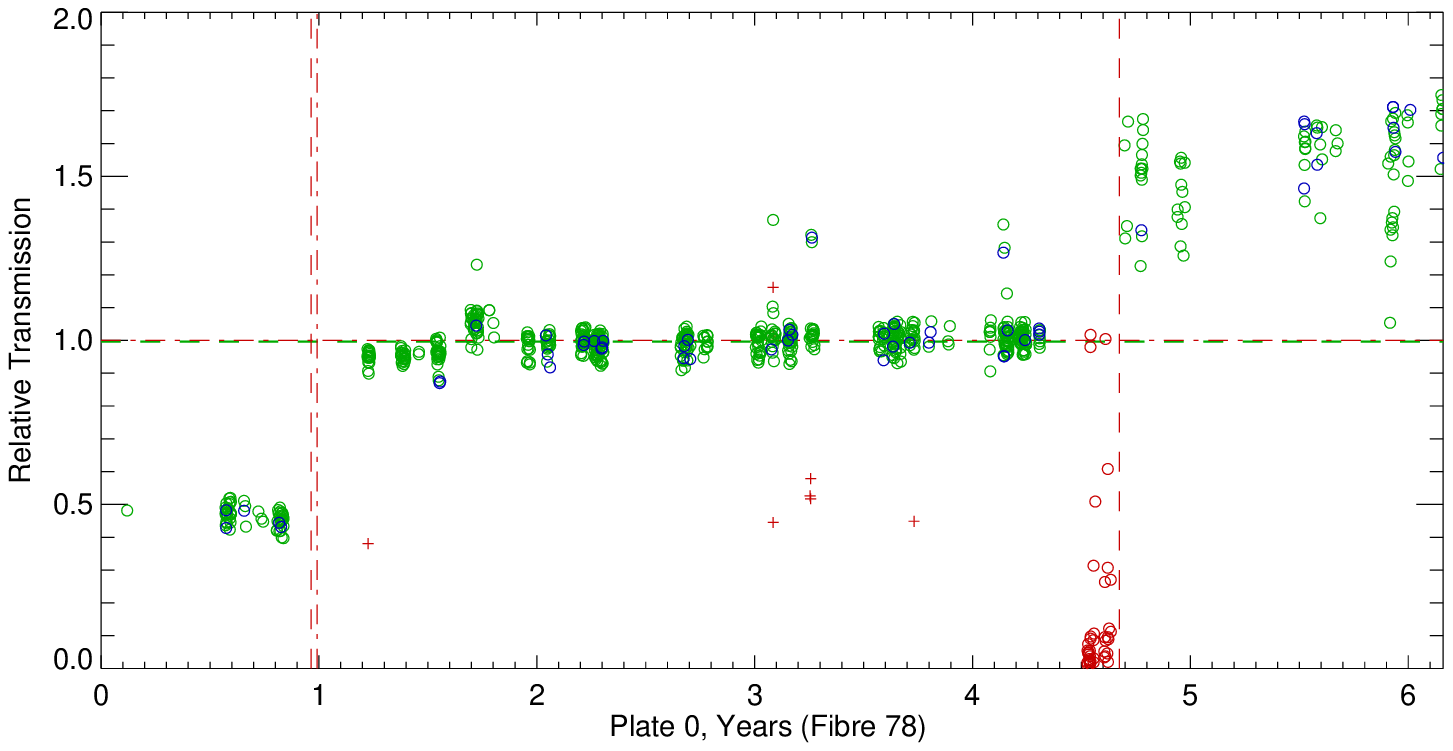,width=8cm}
\epsfig{file=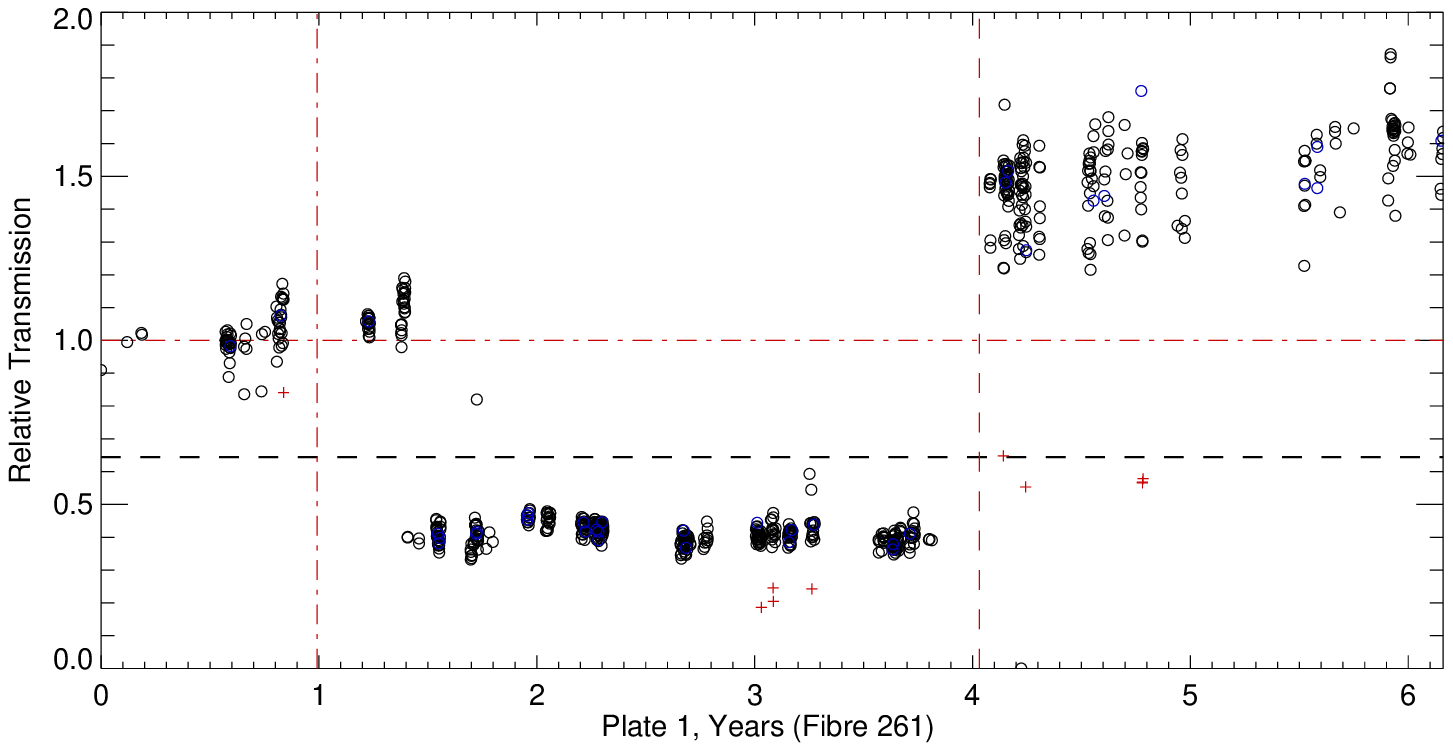,width=8cm}
\caption{\label{FibStep} Step function relative transmission increases are shown for three fibres. Fibre 31 required re-termination, while Fibres 78 and 261 had the fibre re-polished in situ and a new prism fitted. The significant relative transmission gains realised suggest overall performance degradation with time is largely restricted to the fibre button. }
\end{figure}

Many fibres exhibit a significant step-function increase in transmission after repair work, which is of grave concern since the fibre repair database indicates that only 15\% of fibres have had one or more repair events. The relatively small fraction of repairs cannot alone explain the increased scatter in the relative transmission measure over time. Periodic re-termination of fibres which exhibiting poor relative transmission would also likely boost instrument performance. Three such examples are shown in Figure~\ref{FibStep}. The first fibre suffered a breakage requiring re-termination, with the fibre trimmed and mounted in a new steel ferule, before re-polishing and attaching a new prism. The remaining two fibres required only a prism replacement after re-polishing while still in the original magnetic button. The only mechanism we have identified to explain the significant relative transmission gains seen is that of a slow global transmission loss in the full fibre set whose origin is restricted to the fibre button. That significant transmission gains are realised with only replacement of the prism suggests that fibre stress and FRD within the magnetic button are not responsible for the majority of losses. The three most prevalent loss mechanisms are identified as: mechanical misalignment of a prism with respect to the sky, the build up of dirt on the prism optical surface, and, fringing losses at a small prism/fibre air cavity. The first two losses can be avoided with manufacturing quality control and regular assessment and cleaning of the prisms surfaces. New construction methods are under investigation to avoid the fringing losses.

Analysis of the individual life histories reveals three important observations:
\begin{itemize}
\item 40\% of fibres retain constant relative transmission
\item alarmingly, the relative transmission of 20\% of fibres appears systematically to improve over time
\item the majority of light-loss is associated with the fibre/prism interface.
\end{itemize}
Since it has not been possible to conceive of a mechanism by which a fibre's transmission steadily improves over time, one can only surmise that the median relative transmission must decline for all fibres globally.  Fibres whose relative transmission appears to rise are really remaining constant or declining at a reduced rate.

\subsection{A fibre classification based transmission model}
\begin{figure}
\epsfig{file=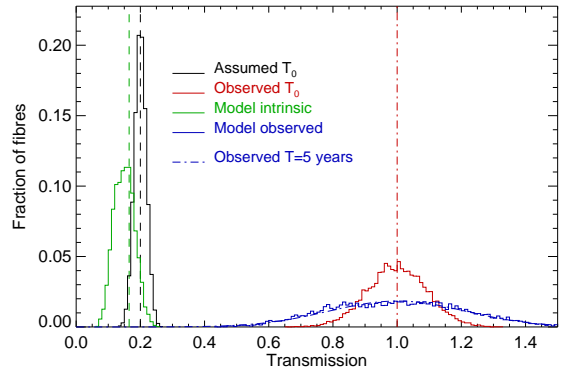,width=8cm}
\caption{\label{model2} An alternate model to that presented in Figure~\ref{Model} is developed for the evolution of fibre relative transmission.  The transmission life history analysis of Figure~\ref{FibreClass} is used to evolve the initial fibre transmission distribution to a distribution that approximates that observed in Figure~\ref{AllRelTranHist} (blue histogram and distribution).  The evolutionary model predicts a higher mean absolute transmission (green histogram) for the fibre distributions over the simplistic stochastic model of of \S\ref{modelsection} due to the wider scatter in relative transmission values when the {\it stable} and {\it rising} (reduced bulk decline) transmission fibre groups are included.}
\end{figure}

The stochastic model from Figure~\ref{Model} can be improved based on the relative percentages of fibres exhibiting each classification and the observed distribution of transmission properties for each classification.  Such a model is shown in Figure~\ref{model2}.  The model recovers the basic properties of the observed relative transmission distribution. The improved accounting for poor fibres and the inclusion of a 25\% population whose transmission appears to rise significantly over time results in a reduced inferred mean fibre transmission loss for the population as a whole.

\section{Conclusions}
The largest single fibre group ($\sim$40\%) shows a stable relative transmission, consistent with a slow decline in absolute transmission (20\% over 4-6--years) contributing the dominate component of the Standard Star sensitivity loss of Figure~\ref{WiggleZstd} and the declining survey completeness reported for the GAMA galaxy redshift survey (\S\ref{Gama}).

A significant ($\sim$17\%) population of fibres present relative transmission rises of $\sim$20\%. These fibres represent a population whose absolute transmission has degraded more slowly than the bulk population - indeed they set the scale for the global decline in transmission in the {\it stable} fibres.

That the majority of {\it stable} fibres exhibit a step-function rise after intervention (prism replacement with or without re-termination) indicates the dominant sources of transmission losses are in the fibre button rather than bulk fibre damage in the extended 38\,m cable or within the fibre retractor system. Loss mechanisms include relativity mundane phenomena such as: poor prisms quality on installation, physical damage to a prism during operation (perhaps due to contact with the 2dF robot), poor prism/button alignment or the buildup of dirt over time, as well as more more involved optical phenomena such as stress induced FRD or interference fringing losses.

There is little evidence for significant increase in bulk fibre transmission degradation over time, either within the portion of the fibre held within the retraction system nor the extended free run. Any such decline in the bulk fibre transmission, or associated losses due to excessive induced FRD, is inconsistent with the available data on the evolution of the relative transmission performance and the individual life-histories of the AAOmega fibre set. This does not preclude the introduction of significant static losses into the fibre feed on initial installation due for example to rapid wear within the retractor before commencement of survey operations.

That significant gains are made with re-polishing and prism replacement alone, rather than via re-termination of the fibre, indicates that the dominant loss mode for almost all fibres is at prism/fibre interface
or within the prism itself due to poor optical quality, poor quality control during button construction, or exposure damage over time.

The significant ($\sim$12\%) step-decline rate and marked differences between functionally identical retractor blocks indicates the clear need for manufacturing quality control and intermediate term monitoring of the fibre feed relative transmission.  The slow-decline profiles of some fibres appears to be due to the same set of manufacturing defects, but with a slower period of degradation mostly related to the prism/fibre/ferrule/button interface. An active program of fibre monitoring and maintenance is to be integrated into routine operations with AAOmega at the AAT as part of preparations for the installation of the new HERMES fibre spectrograph.

\acknowledgments
Thanks are due to Jeremy Allington-Smith for his initial enquires which focused our attention on the need for this analysis, to Will Saunders for his constant enthusiasm for fibre spectroscopy which drives valuable discussion and to Kristin Fiegert for provision of the 2dF fibre button specifications. We thank Ivan Baldry for compiling the GAMA survey completeness statistics, and Sam Richards for a number of invaluable points of discussion. This work would not have been possible without access the intermediate data products from the WiggleZ and GAMA galaxy survey programs.

{\it Facilities:} \facility{AAT (AAOmega)}

\clearpage

\end{document}